\documentstyle[12pt,epsfig]{article}
\def\e{{\rm e}}
\newcommand{\be}{\begin{equation}}
\newcommand{\ee}{\end{equation}}
\newcommand{\bea}{\begin{eqnarray}}
\newcommand{\eea}{\end{eqnarray}}
\newcommand{\nn}{\nonumber}
\begin{document}
\thispagestyle{empty}
\rightline{gr-qc/9701049}
\rightline{BUTP-97/03}
\rightline{January 1997}
\begin{center}{
{\bf NON-ABELIAN SURPRISES IN GRAVITY
\footnote{Lecture presented at the Simi-96,
Tbilisi, Georgia, September 22-28, 1996.
To be published in the Proceedings.}     
}
\vglue 1.5cm
{GEORGE LAVRELASHVILI
\footnote{On leave of absence from Tbilisi
Mathematical Institute, 380093 Tbilisi, Georgia}
}
\vglue 0.5cm
\baselineskip=14pt
{\it Institute for Theoretical Physics, University of Bern}\\
{\it Sidlerstrasse 5, CH-3012 Bern, Switzerland}\\
\vspace{0.3cm}
{ABSTRACT}
} \end{center}
\vglue 0.3cm
{\rightskip=3pc  \leftskip=3pc
\noindent
We present a brief review of some recent results on
non-abelian solitons and black holes in different theories.
\vglue 0.8cm}
\newpage
\section{Introduction} 

\noindent
Since I plan to talk about ``surprises", 
let me first explain how do I understand this word.
Surprise (in physics) is when you get something
what you do not expect with an educated guess. The first surprise
in the field, which I plan to review in the present lecture, 
came in 1988, when Bartnik and McKinnon \cite{BK}
discovered a discrete family of
globally regular, static, spherically symmetric
solutions of the Einstein-Yang-Mills (EYM) theory.
This was not expected from experience in
lower dimensions (EYM theory in (2+1)dimension) \cite{DES},
the Einstein-Maxwell case,  
pure non-abelian theory \cite{COLDES} and pure gravity. 

The important step was to understand that the Bartnik-McKinnon (BK)
solitons are {\it classically} unstable \cite{SZ}.
With guidance from this due it was shown \cite{VG, SW}
that they are not ordinary particles,
but rather gravitational analogues of the electroweak sphalerons
\cite{sphaleron}.
That was beginning of the story.
Now the BK paper gets more then 80 citations in the SLAC database.
It is difficult to review everything, 
so I will bring to your attention selected topics
from my perspective. 

The plan of the lecture is as follows.
I will begin with some general remarks about classical solutions.
The third section starts with a discussion of 
solitons in YMH, EYM, EYMD, and EYMH theories.
Next I will discuss stability analysis  and present a brief 
comparison of different theories. 
In subsection 3.7 I will talk about non-abelian
black holes in the abovementioned theories.
All these are asymptotically flat solutions.

In the Section 4 I will talk about non-asymptotically flat solutions.
Name\-ly, I will present our results on 
the  EYM theory with the cosmological constant.
In my concluding remarks I will speculate about possible 
applications and I will mention some other interesting directions 
which are not covered in this presentation.

The discussion will be rather schematic, but I will try to 
compensate for this by providing a detailed list of references.

\section{General remarks about classical solutions}

A number of interesting classical solutions
with definite physical meaning have been
considered in the literature.

Quite ``famous" are static, finite energy solutions: monopoles 
\cite{monopole} and spha\-le\-rons \cite{sphaleron}. 
Two types of euclidean solutions are known: 
an instanton \cite{instanton} interpolating between vacua 
with different topological numbers and a bounce 
\cite{bounce} describing the decay of a metastable vacuum. 
One can consider as well a namdvilon solution \cite{namdvilon},
which in {\it real time} interpolates between two sphalerons.

We believe that cosmological solutions of the Einstein 
equations describe the evolution of the Universe. 
One should not forget about the most spectacular objects
predicted by General Relativity - black holes. 

I do not plan to discuss here in detail the role and 
meaning of each of these solutions.
The only point I would like to stress is that
some of these solutions have an unstable modes.
So, the instability of a solution does not automatically mean
that the solution is bad (or good).
It might be an essential property, part of the interpretation
like in case of the electroweak sphaleron or bounce.
The sphaleron is unstable since it is ``sitting" at the top of the
barrier separating neighboring vacua.
The bounce has a {\bf single} negative mode \cite{COL2}
which is essential to describe the decay of the metastable state. 

To summarize:
the first step is to find a classical solution,
the next is to understand its role in the (quantum) theory. 
 
\section{Asymptotically flat solutions}
 
\subsection{YM-H sphaleron}

Let us consider YM-H theory in flat space.

The action for the YMH theory has the form
\be
S_{\rm YMH}=
\frac{1}{4\pi}
\int \left( -\, {1\over 4g^2} F^a_{\mu\nu}F^{a\ \mu\nu}
+{(D^\mu\Phi)}^{\dagger} D_\mu\Phi-
\lambda ({\Phi}^{\dagger}\Phi-{v^2\over 2})^2
\right)d^4x , 
\ee
where $F^a_{\mu\nu}$ is the $SU(2)$ gauge field strength,
$F^a_{\mu\nu}=\partial_{\mu} W^a_{\nu}-
\partial_{\nu} W^a_{\mu}+\epsilon^{abc} W^b_{\mu}W^c_{\nu}$,
and $a=1,2,3$ is the $SU(2)$ group index, $\mu,\nu = 0,1,2,3$
are space-time indices.
Covariant derivatives are defined by
$D_\mu\Phi=\partial_\mu\Phi-{i\over 2}\tau^a W^a_\mu\Phi$.

For the gauge field let us take
the usual ``monopole"  ansatz
\be    \label{mansatz}
W_0^a=0, \quad
W_i^a=\epsilon_{aij}{n_j\over r}(1-W(r)) ,
\ee
with  $n_j=x_j/r$ 
and doublet ansatz for the Higgs field
\be
\Phi ={v\over \sqrt{2}}\; H(r)
{0\choose 1}\, .
\ee
The reduced action has the form
\bea  \label{rymhact} 
S^{red}_{YMH}&=&-
\int\Bigl[
\frac{1}{g^2}(W'^2 +{(W^2-1)^2 \over 2r^2}) \nn \\
&+&{r^2\over 2} H'^2+ {H^2 (1+W)^2\over 4}
+\frac{\lambda r^2}{4} (H^2-v^2)^2
\Bigr] dr   
\eea
where a prime denotes $d/dr$.
Out of the gauge coupling constant  $g$,
the Higgs vacuum expectation value $v$,
and the Higgs self coupling constant $\lambda$ one can form two 
mass scales $M_{W}=\frac{1}{2}gv$ and $M_{H}=\sqrt{2\lambda}v$
which are the gauge boson mass and Higgs mass respectively.
 
Solutions with finite energy 
have to interpolate between
$$ W=1 \;,\qquad  H=0\;,$$
at $r\to 0$  and
$$ W=-1\;,\qquad H=v \;$$
for $r\to \infty$.

It was found \cite{sphaleron} that such a solutions $\{ W(r), H(r) \}$
indeed exist. They are called sphalerons.

After suitable rescaling, the energy of the sphaleron can be written as
\be                                        
E_{YMH}=-S^{red}_{YMH}=
\frac{2 M_{W}}{\alpha_{W}} B(M_{H}/M_{W})
\ee
where $\alpha_{W}={g^2}$ is the electroweak ``fine structure" 
constant and the numerical value of the function B varies from  
about 1.5 to 2.7 as $M_{H}$ varies from zero to infinity \cite{sphaleron}.
Since $\alpha_{W}=\frac{1}{29}$  and $M_{W}=80$ Gev, 
the energy of the electroweak sphaleron is of order of 10 Tev.

The main properties of the sphaleron are as follows: 

\noindent
(i) it has finite energy\\
(ii) one can assign fractional topological charge \\
(iii) it is a saddle point of the action \cite{BY} \\
(iv) there are fermion zero modes in background of sphaleron
solutions \cite{RKB} 

Solutions which I discuss in what follows are analogues of sphaleron.
 
If one takes a triplet Higgs, one gets a monopole \cite{monopole}.
Including gravity one obtains gravitating monopoles.
I will not talk about gravitating monopoles.
They are discussed in \cite{BFMmonopole}, \cite{LNW}, \cite{DIMkorea}.
 
\subsection{EYM theory}
 
Let me now describe the discovery of Bartnik and McKinnon.

The action for the EYM theory has the form
\be \label{eymaction}
S_{\rm EYM}=
\frac{1}{4\pi}\int\left(-\frac{1}{4 G}R
-\, {1\over 4g^2} F^a_{\mu\nu}F^{a\ \mu\nu}
\right)\sqrt{-g}\, d^4x 
\ee
where \ $G$ is Newton's constant.

A convenient parametrization for the metric turns out to be
\be \label{schwinterval}
ds^2=S^2(r)N(r)dt^2-{dr^2\over N(r)}-r^2d\Omega^2\;,
\ee
where $d\Omega^2=d\theta^2+\sin^2(\theta)d\varphi^2$
is the line element of the unit sphere.

For the $SU(2)$ YM potential we make the usual (`magnetic')
spherically symmetric ansatz  Eq.~(\ref{mansatz}). 
Substituting this ansatz into the action (\ref{eymaction})
we obtain the reduced action
\be
S^{\rm red}_{\rm EYM}=-\int S\left[{1\over 2G}(N+rN'-1)
+{1\over g^2}\left(N W'^2+{(1-W^2)^2\over2r^2}\right)\right]\,dr\;.
\ee 
Now let me explain what one means when one talks about
Planck mass $M_{Pl}=\sqrt{\hbar c/G}$ scale in relations 
to the BK solutions. This is subtle since the classical  
EYM action does not contain $\hbar$ at all.
For this purpose I will restore $c$ and $\hbar$.
Out of two coupling constants with dimensions 
$[G]=L^3 T^{-2} M^{-1}$ and  $[g]=L^{-1} T^{1/2} M^{-1/2}$
one can form a quantity with dimension of length
$l_{EYM}=\sqrt{G/{c g^2}}$  and mass $m_{EYM}=\sqrt{c/{G g^2}}$.
Introducing the dimensionless ``fine structure constant" 
$\alpha_{g}=\hbar g^2$, the mass scale can be written as 
$m_{EYM}=M_{Pl}/\sqrt{\alpha_{g}}$.  
If we work in dimensionless variables $\hat{r}=r/{l_{EYM}}$,
the energy functional is
\be \label{eymscaling}
E_{EYM}=-S^{red}_{EYM}=
\frac{M_{Pl}}{\sqrt{\alpha_{g}}} S(\hat{r})
\ee 
Now we keep in mind scaling (\ref{eymscaling}), and
put $G=g=1$ and omit a hat over the $r$ in the
resulting field equations 
\bea \label{eymeq}
 (NS W')'&=& S {W(W^2-1)\over r^2} \;, \cr
 N' &=& {1\over r}\left(
 1-N-2\left(N W'^2+{(1-W^2)^2\over2r^2}\right)\right)\;, \cr
 S^{-1}S'&=& \frac{2W'^2}{r} \;.
\eea
The field equations (\ref{eymeq})  have
singular points at $r=0$ and $r=\infty$ as well
as points where $N(r)$ vanishes (horizon!).
The regularity at $r=0$ of a configuration
requires $N(r)=1+O(r^2)$, $W(r)=\pm1+O(r^2)$
and $S(r)=S(0)+O(r^2)$.
$W$ and $-W$ are gauge equivalent and we choose $W(0)=1$. 
Similarly we can assume $S(0)=1$ since a rescaling of $S$
corresponds to a trivial rescaling of the time coordinate.
Inserting a power series expansion into (\ref{eymeq}) one finds
\bea \label{rzero}
W(r)&=& 1-br^2+O(r^4)\;, \cr 
N(r)&=& 1-4b^2r^2+O(r^4)\;, \cr
S(r)&=& 1+4b^2r^2+O(r^4)\;, 
\eea
where $b$ is an arbitrary parameter.

Similarly assuming a power series expansion in $1\over r$ at $r=\infty$
for asymptotically flat solutions, one finds
$\lim\limits_{r\to\infty}W(r)=\{\pm 1,0\}$.
It turns out that $W(\infty)=0$ cannot occur for 
globally regular solutions. For the remaining cases one finds
\bea \label{rinfty}
 W(r)&= \pm \left(1-{c\over r}+O\left({1\over r^2}\right)\right)\;, \cr
 N(r)&= 1-{2m\over r}+O\left({1\over r^4}\right)\;, \cr
 S(r)&= S_\infty\left(1+O\left({1\over r^4}\right)\right)\;,
\eea
where again $c, m$  and $S_\infty$ are arbitrary
parameters and have to be determined from numerical calculations.

It was found \cite{BK} that equations (\ref{eymeq}) admit
a discrete sequence of finite-energy solutions
$\{W_n, N_n, S_n \}$ which interpolate between
the asymptotic behaviors (\ref{rzero}) for $r \to 0$
and (\ref{rinfty}) for $r \to \infty $.
Solutions can be labeled by an integer $n$
the number of zeroes of the gauge amplitude $W$. 
The energy (mass) of the solutions is given
\be
M^{(n)}_{EYM}= \frac{M_{Pl}}{\sqrt{\alpha_{g}}} m_n
\ee
where $m_n$ takes values between 0.82 and 1.0 when $n$ varies from
one to infinity \cite{BK, BFM}.

\subsection{EYMD theory}

There are many reasons to believe that there is a dilaton field. 
 
Introducing a dilaton field we naturally
obtain a EYMD theory with the action
\be  \label{eymdaction}
S_{\rm EYMD}=
\frac{1}{4\pi}\int\left(-\, \frac{1}{4 G}R+
\frac{1}{2}(\partial\varphi)^2-
{\e^{2\kappa\varphi}\over 4g^2}F^2\right)\sqrt{-g}\, d^4x ,
\ee
where $\kappa$ denotes the dilatonic coupling constant.
After proper rescaling this theory depends on a dimensionless
parameter $\gamma=\frac{\kappa}{g\sqrt{G}}$.
The model (\ref{eymdaction}) was analyzed \cite{LM2, BIZ2, DG, TM} 
and for any value of the dilaton coupling constant a discrete 
family of globally regular solutions of finite mass was found. 

A few remarks are in order.
 
In the limit $\gamma\to 0$ one gets the EYM theory studied in
\cite{BK}. 

The value $\gamma=1$ corresponds to a model obtained from
heterotic string theory.
A very special situation occurs for this value of
the dilaton coupling constant $\gamma=1$.
It was found \cite{LM2} that for the $n=1$ solution
the parameter $b$ is a {\bf rational} number, $b =\frac{1}{6}$.
Another regularity found numerically is that an asymptotic coefficient
$c$ in the equation analogous to the Eq.(\ref{rinfty})
is related to the mass of the solution, $c=2m$.
In addition one has one Bogomol'nyi type equation
\be
g_{00}=e^{2\gamma\phi}.
\ee
We think these are arguments indicating that the lowest lying
($n=1$) regular solution in the EYMD theory may be obtained 
in a closed form similarly to the ``stringy instanton''
and the ``stringy monopole'' \cite{instmon}.

In the limiting case $\gamma\to\infty$ 
one  obtains the YM-dilaton theory in
flat space  \cite{LM1, BIZ1}. The YMD system is quit
different from  EYM, but surprisingly one obtains again
tower of solutions with properties similar to
the BK solutions.

I would stress as a main difference with the EYM case the 
existence of magnetically charged limiting solution with 
infinitely many zeroes \cite{LM1, BIZ1}. 

Another important point is that the masses of YMD solutions
(as well as EYMD for strong coupling $\gamma >>1$) are
inverse proportional to the dilatonic coupling constant 
\be
M^{(n)}_{YMD}= \frac{M_{Pl}}{\gamma\sqrt{\alpha_{g}}} m_n
\ee 
 
\subsection{EYMH theory}

One can consider the combined EYMH theory.
This theory was analyzed \cite{GMN} and the results 
confirm what one can expect from EYM and YMH cases. 

The theory has four coupling constants $G, g, v$ and $\lambda$
or three mass scales $m_{EYM}, M_{W},$ and $M_{H}$.
After a suitable rescaling the EYMH theory depends on two dimensionless 
ratios $\alpha=M_{W}/(\alpha_{g} m_{EYM})$
and $\beta=M_{H}/M_{W}$.
For any given $\beta$ one gets 1-parameter family of solutions.
It is important that for small  $\alpha\neq 0$ the solutions bifurcate:
there are two different solutions with any given number of nodes. 
In fact there are two types of nodes: one is a BK type, with 
typical size $l_{EYM}$ and the other a sphaleron type with $l_{W}=1/M_{W}$.
With increasing $\alpha$, the sphaleron type node moves inwards and
at some value $\alpha_{max}$ the two solutions merge and cease to exist
for bigger values of $\alpha$ \cite{GMN, LM3, DIMkorea, MW, DENSING}.

\subsection{Stability analysis}

In order to analyze the stability of the above discussed solutions 
we have to consider the spectrum of (harmonically) time dependent 
perturbations in the background of a given solution.
The existence of imaginary frequencies in the spectrum of the
linearized equations (leading to an exponential growth of the
initially small perturbations in time) 
indicates the instability of the background solution. 

One should be careful in stability analysis. The
point is that the most general spherically symmetric
ansatz for the {\sl SU(2)\/} Yang-Mills field $W_\mu^a$ can be written
(in the Abelian gauge) as \cite{ANSATZ} 
\bea
W_t^a&=(0,0,A_0)\,,\qquad  W_\theta^a&=(\phi_1,\phi_2,0)\, \nn \\
W_r^a&=(0,0,A_1)\,, \qquad
W_\varphi^a&=(-\phi_2 \sin\theta,\phi_1\sin\theta,\cos\theta)\,.
\label{gauge}
\eea
and contains four functions, whereas sphaleron like-solutions 
lie in a special ansatz (\ref{mansatz}) with only one gauge amplitude.
One can consider perturbations within and outside of the ansatz.
It turns out they form two different sectors \cite{LM3, BBMSV, VGodd}.
Unstable modes of the first type we call gravitational
instabilities since they have no analogue for the flat space sphalerons.
Instabilities of the second type we call sphaleron instabilities,
because they have the same nature as the instability found for the 
electroweak sphaleron \cite{BY}.

The gravitational sector for the BK solutions
was investigated already in 1990 \cite{SZ}
and $n$ unstable modes were found for the $\rm n^{th}$ solution.

Recently the sphaleron sector was carefully analyzed
\cite{LM3, VBLS} and an extra $n$ negative modes
for the $n^{th}$ solutions was found. 
Altogether the $\rm n^{th}$ solution has  $2n$ negative modes
within the most general spherically symmetric ansatz.
This is true for the
EYM \cite{SZ, LM3, VBLS},
EYMD \cite{ASCH, DENSING}, 
EYMH \cite{BBMSV, MW}
solutions.

The numerical values
for the energies $E=\omega^2$ of the negative modes of
the first three BK solutions in the gravitational respectively
sphaleron sectors are shown in Tables 1 and 2 \cite{LM3}:
 
\begin{center}
\begin{tabular}{|l|l|l|}  \hline
$n=1$        &$n=2$        &$n=3$          \\ \hline
$E_1=-0.0525$&$E_1=-0.0410 $&$E_1=-0.0339 $ \\
             &$E_2=-0.0078 $&$E_2=-0.0045 $ \\
             &              &$E_3=-0.0006 $ \\  \hline
\end{tabular}
\vglue 0.4cm
Tab 1. Bound state energies for the $n=1,2,3$  BK solutions, \\
(gravitational sector).
\end{center}
 
\begin{center}
\begin{tabular}{|l|l|l|l|}  \hline
$n=1$        &$n=2$        &$n=3$         \\ \hline
$E_1=-0.0619$&$E_1=-0.0360$&$E_1=-0.0346$ \\
             &$E_2=-0.0105$&$E_2=-0.0037$ \\
             &             &$E_3=-0.0009$ \\ \hline
\end{tabular}
\vglue 0.4cm
Tab 2. Bound state energies for the $n=1,2,3$  BK solutions,\\
(sphaleron sector).
\end{center} 

\subsection{Comparison of different theories} 

As mentioned earlier, there are no static solutions in
the pure YM theory in $(3+1)$ dimension \cite{COLDES}. 
The reasons are that pure YM theory is repulsive and has no scale.
In order to have solutions with finite energy one needs some extra 
field which breaks scale invariance and provides an 
attraction which compensates the YM repulsion. In the case of the 
electroweak sphaleron, this job is done by a Higgs field.
As we discussed earlier, it can be done by gravity \cite{BK}
and by the dilaton field \cite{LM1, BIZ1}. 

The YM configuration is essentially the same and the EYM and 
the YMD solutions share the main properties of the sphaleron.
Namely, they have finite energy, fractional charge \cite{MOSSW, GLzero} 
and there are fermion zero modes in the background of these 
solutions \cite{GS, GLzero}.

The situation is summarized in the Tab.3.
 
\begin{center}
\begin{tabular}{|l|c|c|c|c|}  \hline
Properties of  solitons & YMH   & EYM                 & (E)YMD \\
in different theories   &              &             &          \\ \hline
                        &              &             &          \\ 
Finite energy           &$2 B(M_{H}/M_{W}) {M_{W}}/{\alpha_{W}} $
&$M_{Pl}/\sqrt{\alpha_{g}}$ &$M_{Pl}/\gamma\sqrt{\alpha_{g}}$ \\ 
Fractional charge   & $1/2$ &  $1/2$      & $1/2$           \\
Negative mode(s)    &   yes & yes (2n!)   & yes (2n!)        \\
Existence of fermion&       yes    &     yes     &  yes      \\
zero modes          &              &             &          \\ \hline 
\end{tabular} 
\vglue 0.4cm
Tab 3. Main properties of solitons in different theories.
\end{center}

\subsection{Sphaleron black holes}

There are corresponding black holes
in EYM, EYMD, EYMH theories \cite{bholes, LM2, GMN}.
This is in a way a non-trivial fact since not all
classical lumps allow a horizon \cite{KT}. 

One of the most interesting results of this activity is
finding of a violation of the No-Hair conjecture.
There was a widespread belief that black holes are completely
characterized by their ``quantum numbers" seen from infinity:
mass, electric and magnetic charges, angular momentum.
J.A. Wheeler put this statement as ``black holes have no hairs".
There are  No-Hair theorems for theories with scalar fields
\cite{SUD, BECK} and Maxwell field \cite{HEUSLER, CHRU}.

The non-abelian case shows that this conjecture is not valid.
In the EYM and EYMD theory there are non-abelian black holes 
which have the same quantum numbers as Schwarzschild hole but 
are different from Schwarzschild!

Very unfortunately the sphaleron black holes 
share the instability of the globally regular solutions.
Magnetically charged non-abelian black holes 
\cite{BFMmonopole, LNW, DIMkorea} provide examples
for stable counter examples for the No-Hair conjecture.
 
\section{Non-asymptotically flat solutions. 
EYM theory with the cosmological constant}

In this section I will briefly touch the situation in which
the solutions are non-asymptotically flat 
\cite{MAEDA, VSLHB, BHLSV}.

The action for the EYM theory with the cosmological constant
$\Lambda$ has the form:
\be \label{eymlaction}
S_{EYM_{\Lambda}}=
\frac{1}{4\pi}\int\Bigl(-\frac{1}{4 G}(R+2\Lambda)
- {1\over 4g^2}F^2\Bigr)\sqrt{-g}d^4x . 
\ee
It turns out that some of the solutions of the model (\ref{eymlaction})
have critical points, and Schwarzschild - like
coordinates (\ref{schwinterval}) are no more suitable.
A convenient system of coordinates in this case turns out to be 
\be\label{genint} 
ds^2=Q^2(\rho)dt^2-\frac{d\rho^2}{Q^2(\rho)}-r^2(\rho) d\Omega^2\;.
\ee 
The EYM${_{\Lambda}}$ theory was analyzed and a whole bunch of
solutions was found \cite{VSLHB}. 
Close to $r=0$ they can be parameterized as in BK case 
by the shooting parameter $b$. Fig.1 shows the 
shooting parameter $b$ versus $\Lambda$ for different 
solutions.  For small $\Lambda$ one obtains solutions which generalize the 
BK solitons, but now are surrounded by a cosmological horizon and 
asymptotically approach the de Sitter geometry.
Increasing $\Lambda$ one obtains  solutions which have
both equator (critical point where $dr/d\rho =0$) and horizon.
With increasing cosmological constant the horizon shrinks to zero size
and the curves in Fig.1 end with regular solutions which are
topological spheres.
 
In fact one can integrate the system  of equations
and obtain $n=1$ regular solution in closed form:
\be \label{HDH}
Q=1 \ , \
r=\sqrt{2} sin({\rho\over \sqrt{2}}) \ , \
W=cos({\rho\over \sqrt{2}}) \ , \  
\Lambda ={3\over 4}
\ee 
A stability  analysis of these solutions  \cite{BHLSV} is much more
involved than in asymptotically flat case. The main source of
difficulties is the existence of a critical point in the background 
solution. It was shown that pulsation equations can be brought to a form
convenient for numerical analysis  \cite{BHLSV}. The conclusion is again 
that the $n^{th}$ solution has $2n$ unstable modes. Since the system of 
pulsation equations in the gravitational sector is in a non-standard form, 
textbook theorems are not applicable, and mathematical questions about 
spectrum of bound states (that it is real and discrete) are still open.  

\section{Concluding remarks}

To conclude, let me list what I think are
main non-abelian surprises:

-Existence of nontrivial globally regular solutions
in the EYM and YMD theory. 

-Violation of the No-Hair conjecture in non-abelian
case.

-Interplay between gravity and YM which
leads to the solution Eq.~(\ref{HDH}).
  
-EYMD theory, string case, $n=1, \gamma=1$ analytic solution (?).
\footnote{ This case I would rather call as  an ``anti-surprise" 
according my definition. There was hope (and strong indications!)
that one can find solution in a closed form 
but till now nobody was able to find it.} 
 
So far a whole zoo of solutions has been found.
A natural question arises: what might be a
possible application for these non-abelian ``animals".
Let me mention a few ideas:

-Fermion number non-conservation.\\
One finds fermion zero modes in the background of EYM and YMD
solitons leading to fermion number non-conservation.
Can one hope about a possible generation of the baryon asymmetry?
I think the answer is No. The masses are too high
and they cannot ``compete" with the electroweak sphaleron.
Another problem is that the gravitational and dilatonic sphalerons
have {\bf even} number of negative modes. The
EYM example shows (Tab.1 and Tab.2) that instabilities
in the different sectors have comparable energies and therefore
one cannot neglect gravitational instabilities and apply formulas
which are designed for the case with a single negative mode.
This question needs further investigation.
 
-Counterexamples for a No-Hair conjecture.\\
This is very important observation. 
There are some speculations about black holes as elementary particles.
If the No-Hair conjecture had been true, there would have been
no chance to assign lepton or baryon number to a hole (particle).  
Non-abelian black holes show that this is not totally excluded.

-Cosmological applications (?)
 
-Interpretation as an instanton. \\
Static solutions can be considered as an instantons in lower dimensional 
space. This might be a direction in which to think as well. 

There are many interesting developments, 
which I could not cover in this presentation.
Let me mention that the existence proof for EYM solitons 
and black holes has been given \cite{existence, BFM}, the
dynamical evolution of the perturbed BK solutions was studied
\cite{ZS, CHOPBIZ} and the relation to the critical phenomenon 
in black hole collapse was investigated \cite{CHOPBIZ}.
Also axially symmetric solutions of the EYMD theory have been
constructed \cite{KK} recently.

\section{Acknowledgments} 

This work was supported by the Tomalla Foundation. 

\newpage
\begin{figure}
\epsfig{file=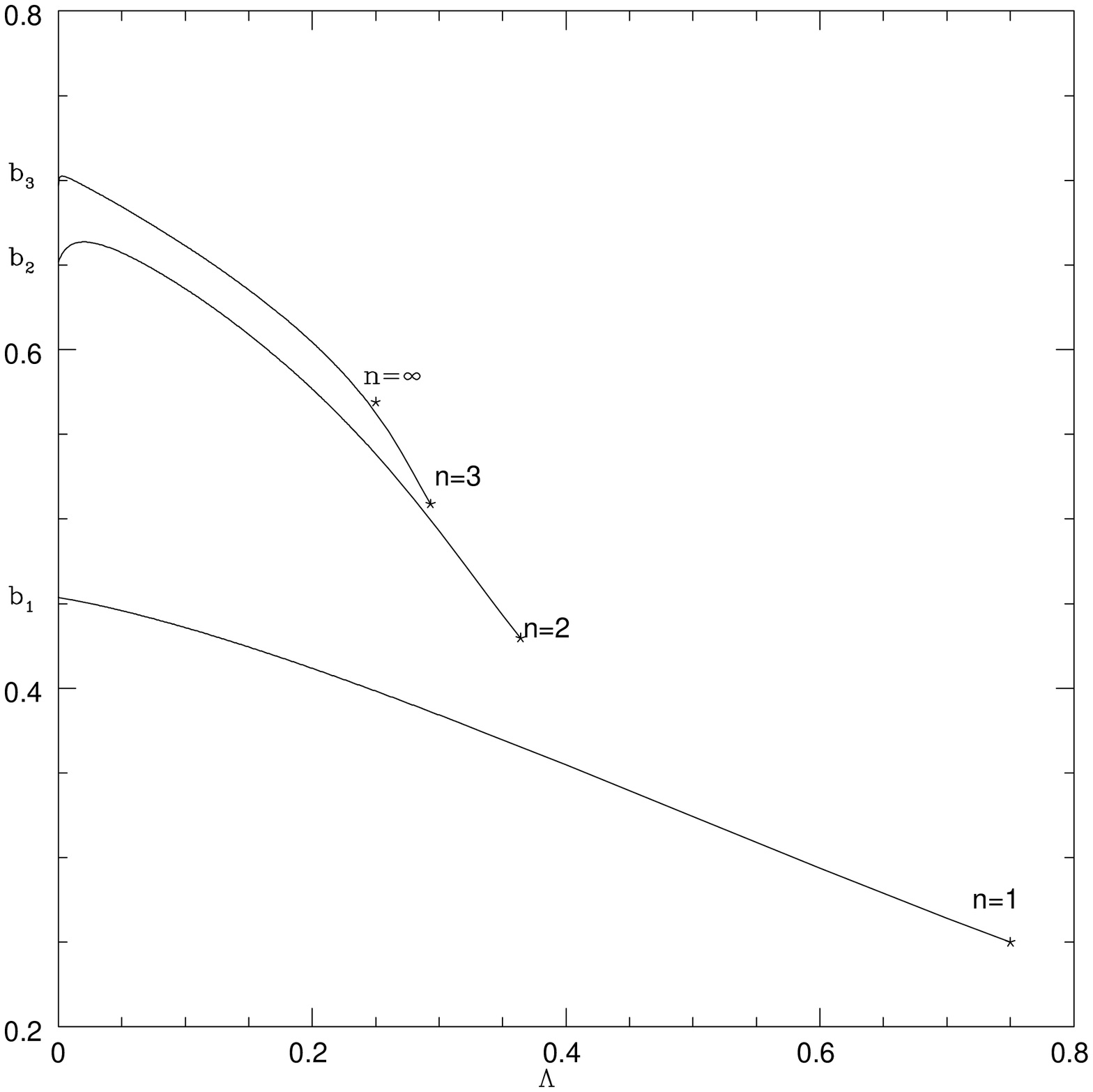,  width=0.9\hsize 
}  
\caption[]{The shooting parameter $b$ versus cosmological constant
$\Lambda$ for different solutions of the EYM${_\Lambda}$ theory.}
\end{figure}
\end{document}